\documentclass[]{spie}  

\usepackage{amsmath,amsfonts,amssymb}
\usepackage{graphicx}
\usepackage[colorlinks=true, allcolors=blue]{hyperref}

\title{The Infrared Imaging Spectrograph (IRIS) for TMT: Prototyping of cryogenic compatible stage for the Imager}

\author{Fumihiro Uraguchi}
\author{Yoshiyuki Obuchi}
\author{Bungo Ikenoue}
\author{Sakae Saito}
\author{\\Ryuji Suzuki}
\author{Yutaka Hayano}
\affil{National Astronomical Observatory of Japan, 2-21-1 Osawa, Tokyo, Japan}

\authorinfo{Further author information: (Send correspondence to F.U.)\\F.U.: E-mail: fumihiro.uraguchi@nao.ac.jp, Telephone: +81 (0)422 34 3861}

\begin{document} 
\maketitle

\begin{abstract}
The IRIS Imager requires opt-mechanical stages which are operable under vacuum and cryogenic environment. Also the stage for the IRIS Imager is required to survive for 10 years without maintenance. To achieve 
these requirements, we decided prototyping of a two axis stage with 80 mm clear aperture. The prototype was designed as a double-deck stage, upper rotary stage and lower linear stage. Most of components are selected to take advantage of heritage from existing astronomical instruments. In contrast, mechanical components with lubricants such as bearings, linear motion guides and ball screws were modified to survive cryogenic environment. The performance proving test was carried out to evaluate errors such as wobbling, rotary and linear positioning error. 
We achieved 0.002 $\rm deg_{rms}$ wobbling, 0.08 $\rm deg_{0-p}$ rotational positioning error and 0.07 $\rm mm_{0-p}$ translational positioning error.
Also durability test under anticipated load condition has been conducted. In this article, we report the detail of mechanical design, fabrication, performance and durability of the prototype.
\end{abstract}


\section{INTRODUCTION}
\label{sec:intro}  

The Infrared Imaging Spectrometer (IRIS) for TMT will be capable of diffraction-limited imaging and integral-field spectroscopy in the wavelength range of 0.84 to 2.4 $\mu$m. The IRIS is international collaborative project consisting of three partner countries including
the United States, Canada, China and Japan.
The design of the IRIS Imager is in progress at the Advanced Technology Center (ATC), National Astronomical Observatory of Japan (NAOJ) that is the research center for advanced instrumentation and technology development. Contribution of ATC 
covers wide range of the design, construction, and performance evaluation of the IRIS Imager. 

One of challenges of the IRIS Imager is multi-axis motion control of optical components. This is required in order to change observation modes and minimize the wave front error due to motion of the field. The motion control can be provided by an opt-mechanical stage which is operable under vacuum and cryogenic environment. In addition, the stages for the IRIS Imager are required to survive for 10 years without maintenance. One of the stages is for the cold stop that is located at a pupil and reduces background thermal noise. First of all, we assumed that a stage for the cold stop carries an 80 mm diameter pupil mask and requires one rotational axis and two translational axes to align and retract the pupil mask. We noticed that it was hard to obtain a three axis stage with 80 mm clear aperture from commercial market and then decided to start prototyping. 

\section{PROTOTYPING}
\label{sec:prototyping}

\subsection{Objective}

The objective of prototyping is to demonstrate adequate performance 
of opto-mechanical stage, which meets requirement. However, the final requirement was not defined at the moment when the prototyping started. 
The immediate development goal
of the prototype in tab.~\ref{tab:goal} is commonly probable for the stages of the IRIS Imager. 
Axes of the prototype stage were
limited to one rotation and one translation 
motion due to 
the space of inside of cryostat of test environment. 

\begin{table}[ht]
\caption{Development goal of the prototype. This immediate goal is commonly probable for the stages of the IRIS Imager} 
\label{tab:goal}
\begin{center}       
\begin{tabular}{llcl} 
\hline
\rule[-1ex]{0pt}{3.5ex}  Range & (Rotation)      & $\pm$ 360 & deg  \\
\rule[-1ex]{0pt}{3.5ex}        & (Linear motion) & $\pm$ 3   & mm   \\
\hline
\rule[-1ex]{0pt}{3.5ex}  Accuracy & (Rotation, wobble) & 0.05 & $\rm deg_{0-p} (1\sigma)$  \\
\rule[-1ex]{0pt}{3.5ex}        & (Rotation, drive)     & 0.05 & $\rm deg_{0-p} (1\sigma)$   \\
\rule[-1ex]{0pt}{3.5ex}        & (Linear motion)       & 0.05 & $\rm mm_{0-p} (1\sigma)$   \\
\hline
\rule[-1ex]{0pt}{3.5ex}  Lifetime & (Rotation)      & 2.2 $\times 10^6$ & deg  \\
\rule[-1ex]{0pt}{3.5ex}           & (Linear motion) & 4.5 $\times 10^6$ & mm   \\
\hline
\rule[-1ex]{0pt}{3.5ex}  Operation temperature &  & $\leq$ 100 & K  \\
\rule[-1ex]{0pt}{3.5ex}  Clear aperture            &  & 80 & mm   \\
\hline
\end{tabular}
\end{center}
\end{table}

\subsection{Mechanical Design}
\label{sec:mechanicaldesign}

\begin{figure} [ht]
   \begin{center}
   \begin{tabular}{c} 
   \includegraphics[width=15cm]{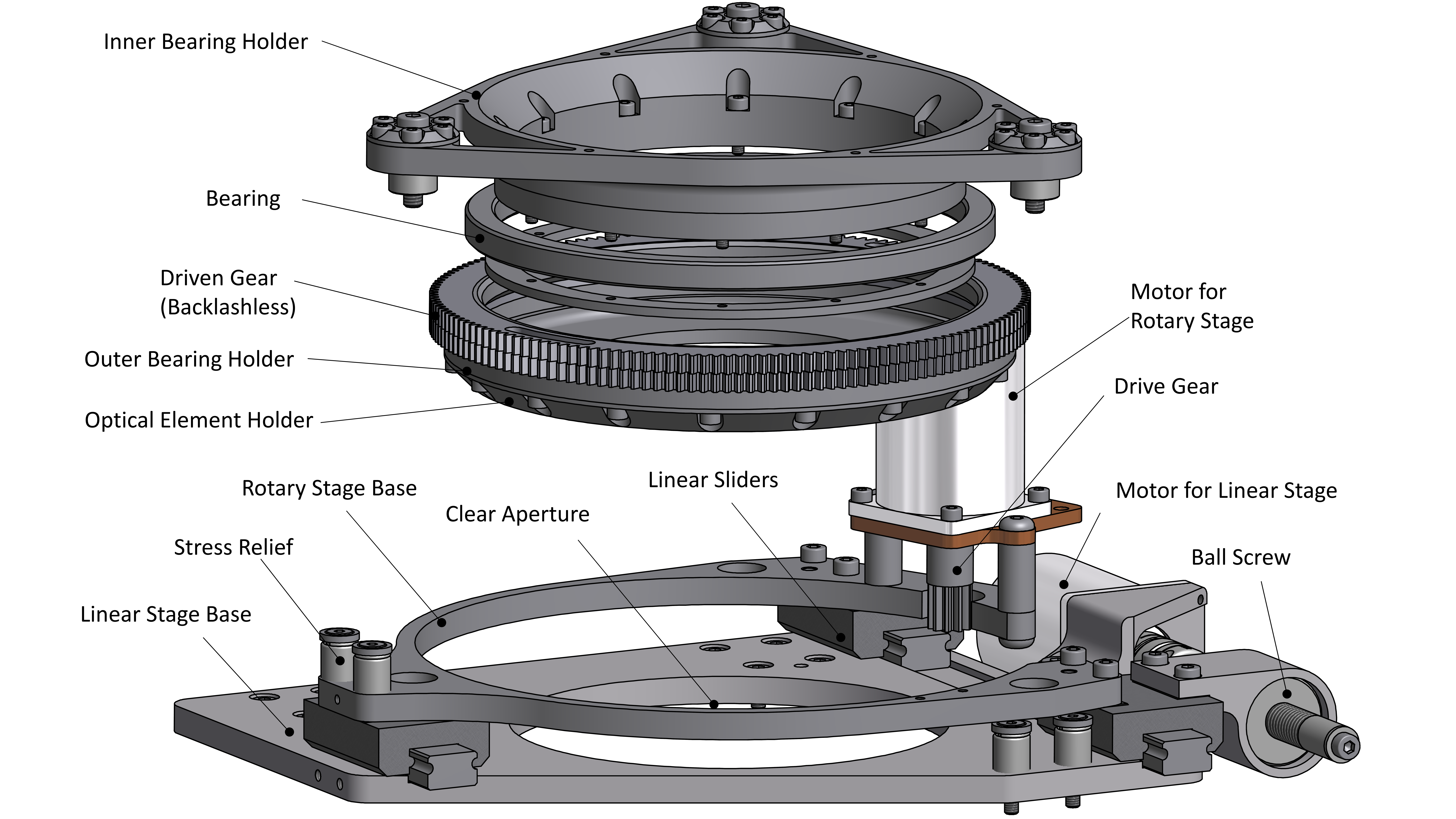}
   \end{tabular}
   \end{center}
   \caption[exploded] 
   { \label{fig:exploded} 
Exploded view of the prototype. Upper part and lower part of the figure describes the rotary stage and the linear stage, respectively.}
   \end{figure}

The prototype was designed as a double-deck stage, upper rotary stage and lower linear stage. Fig.~\ref{fig:exploded} describes components of the stage. The stage is made of stainless steel with consideration of coefficient of thermal expansion (CTE) of rotary bearing material, however structural parts such as the linear stage base are made of aluminum in terms of weight reduction. Connecting parts between different materials equip stress relief mechanisms to absorb CTE difference between stainless steel and aluminum. 

\begin{figure} [ht]
   \begin{center}
   \begin{tabular}{c} 
   \includegraphics[width=15cm]{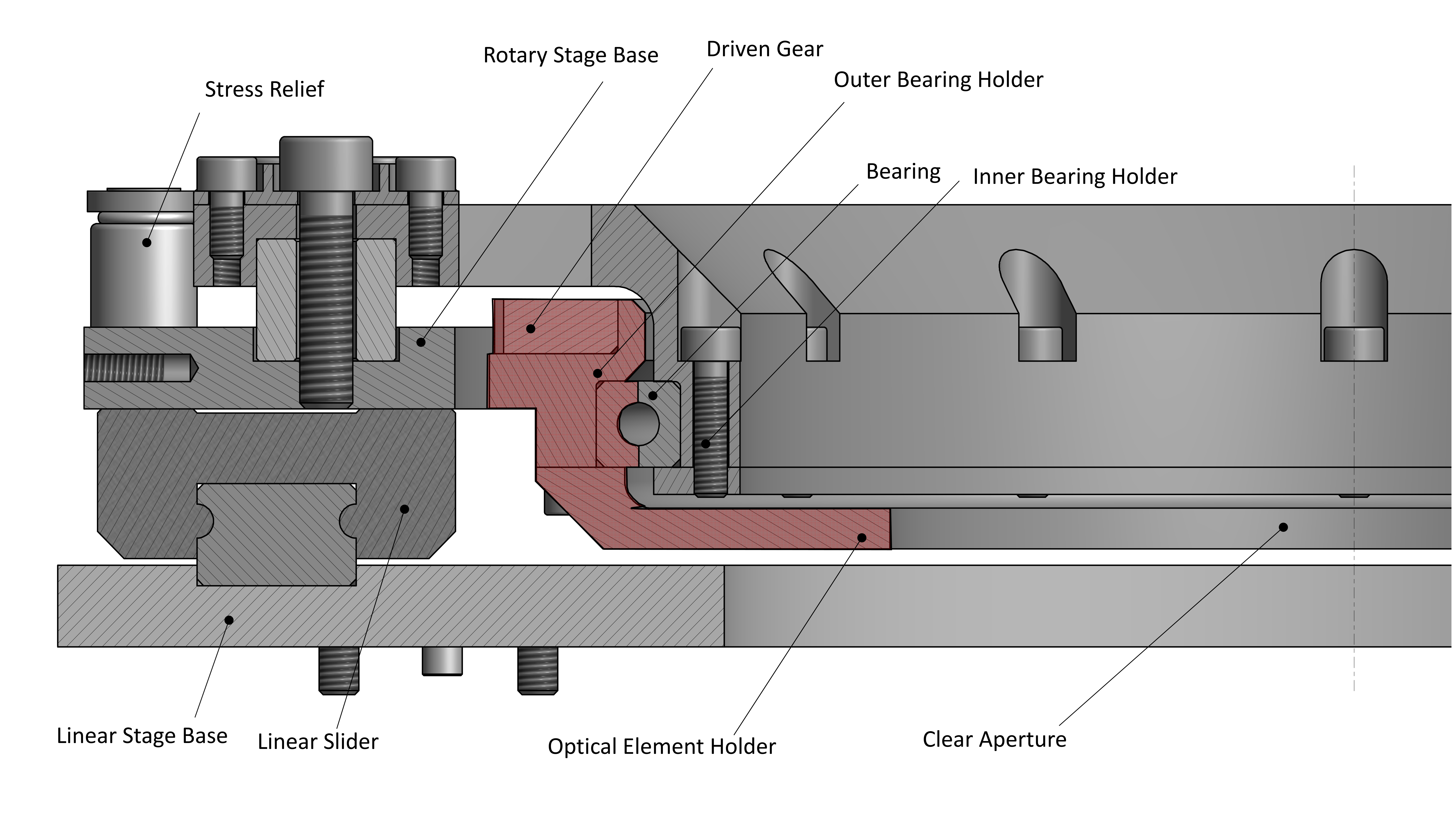}
   \end{tabular}
   \end{center}
   \caption[crosssection] 
   { \label{fig:crosssection} 
Cross section view of the prototype. The optical element will be attached on the outer part (colored in red) of the rotary stage. Stress relief mechanism absorbs CTE difference between the rotary stage base and the linear stage base.}
   \end{figure}

The rotary stage consists of the inner part and the outer part that is colored in red in fig.~\ref{fig:crosssection}. The optical elements will be attached on the outer part of the rotary stage in consideration of difference of thermal contraction between the inner and outer bearing holders.
The rotation part of the bearing is poor thermal conductivity due to small contact area between balls and races. Moreover, it is hard to implement cooling path to the rotation part. These cause temperature gradient between the inner and the outer race of the bearing. In principle, the inner race must be cooler than the outer race otherwise the outer race will squeeze bearing balls due to difference of thermal contraction. Another concern is that heat from
the motor for the rotary stage
will generate temperature gradient on the rotary stage base and 
would cause uneven internal clearance on the bearing. It is anticipated that insufficient internal clearance will generate unexpected stress and defects on balls and tracks in the bearing ultimately. To 
avoid the failure of the bearing, the rotary stage base was designed as not ordinary shape but horseshoe shape. The horseshoe shape allows a part close to the motor mount on the end of the horseshoe expands locally. Thanks to this design, strain due to local thermal expansion does not propagate to the inner bearing holder so that the bearing holder and the bearing can maintain proper shape.

The rotary stage is driven by 160 teeth driven gear and 10 teeth drive gear on the motor. The driven gear consists of two identical layers that are connected by springs and these gears lightly apply a preload to the drive gear in order to eliminate mechanical backlash.

Lubrication for mechanical components such as bearings, linear motion guides and ball screws was 
reassessed
in order to survive cryogenic environment. In such an environment, solid lubricants such as molybdenum disulfide ($\rm MoS_2$) coating have been commonly used. However, conventional solid lubricants have lower load capacity and shorter lifetime compared to standard products. Another concern is that balls and tracks may be stuck due to cold welding once lubricants are depleted.    

For the rotary motion, we selected SA040XT4 stainless steel REALI-SLIM bearing from Kaydon Corporation as a base product. Standard stainless balls were replaced with precision silicon nitride ($\rm Si_3N_4$) balls from Tsubaki 
Nakashima co., ltd. Because this hybrid bearing can eliminate chance of cold welding in principle, it is suggested that the hybrid bearing can be used without lubricants.  The load capacity of the hybrid bearing is expected to be comparable to the standard products\cite{Tanimoto2000}, however performance without lubrication and failure modes are not well known. We expected some indication could be obtained by the performance proving test which is described in sec.~\ref{sec:performance}. 

For the linear motion, we selected set of a linear motion guide and a ball screw as a mechanism. The performance of these components are proved by existing astronomical instruments such as COMICS for Subaru Telescope. We selected BS0802A(2) precision ball screw from KSS Co.Ltd. and RSR12VSE full-ball type LM guide from THK Co.Ltd.. For both components, standard stainless balls were replaced with ones processed with MoST\texttrademark coating\cite{TeerCoatingsWebsite}. The MoST\texttrademark coating is $\rm MoS_2$/metal composite coating developed by Teer Coatings Ltd. According to the website of Teer Coatings Ltd., MoST\texttrademark coatings are harder and more wear resistant than traditional $\rm MoS_2$ coatings, yet still retain the low friction characteristics of $\rm MoS_2$.

On the modification of mechanical components, we had one common design philosophy that was to keep a positive internal clearance in order to avoid an interference fit which will cause unanticipated friction and cold welding in a case of uneven thermal contraction. This design will guarantee smooth motion in cryogenic environment, however this will result negative side effects such as increase of random errors and lost motions. This effect can be identified by the performance proving test.

\begin{figure} [ht]
   \begin{center}
   \begin{tabular}{c} 
   \includegraphics[width=13cm]{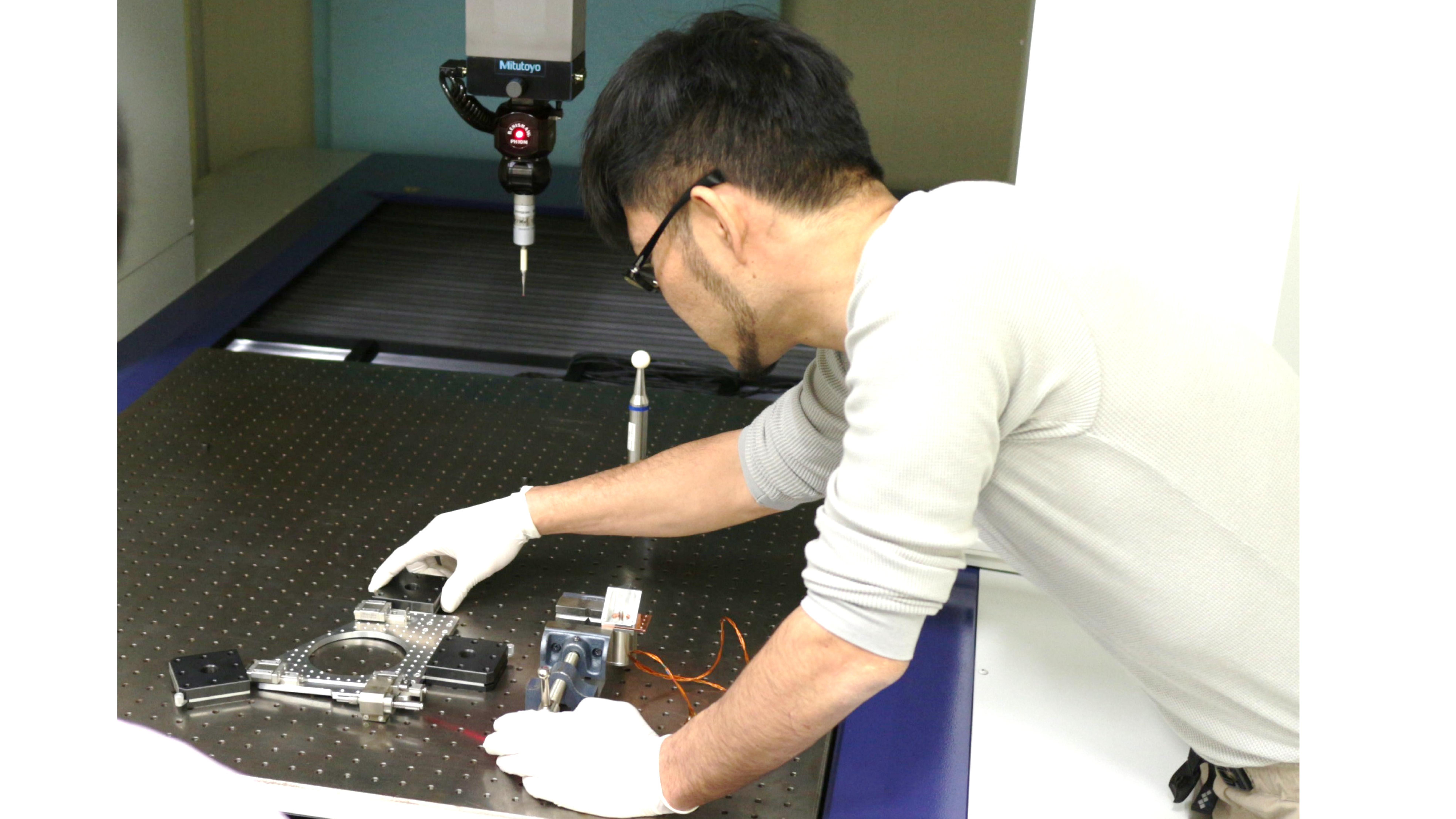}
   \end{tabular}
   \end{center}
   \caption[cmm] 
   { \label{fig:cmm} 
Alignment of drive axes of the linear stage with 3D coordinate measuring machine at ATC.}
   \end{figure}

The stage parts such as drive mechanisms including 160 teeth driven gear for the rotary stage were fabricated at Mechanical Engineering Shop of ATC, NAOJ. The prototype was also assembled at ATC, NAOJ including alignment of drive axes of both the linear sliders and the ball screw by using a 3D coordinate measuring machine as shown in fig~\ref{fig:cmm}. Finished dimension of the prototype is approximately 220 mm (W) $\times$ 180 mm (D) $\times$ 95 mm (H), and the height of the aperture part is 32 mm. Weight is 1.92 kg without the motor and the ball screw for the linear stage. Fig.~\ref{fig:prototype}(a) shows finished prototype. 

\subsection{Motion Control}

\begin{figure} [ht]
   \begin{center}
   \begin{tabular}{c} 
   \includegraphics[width=15cm]{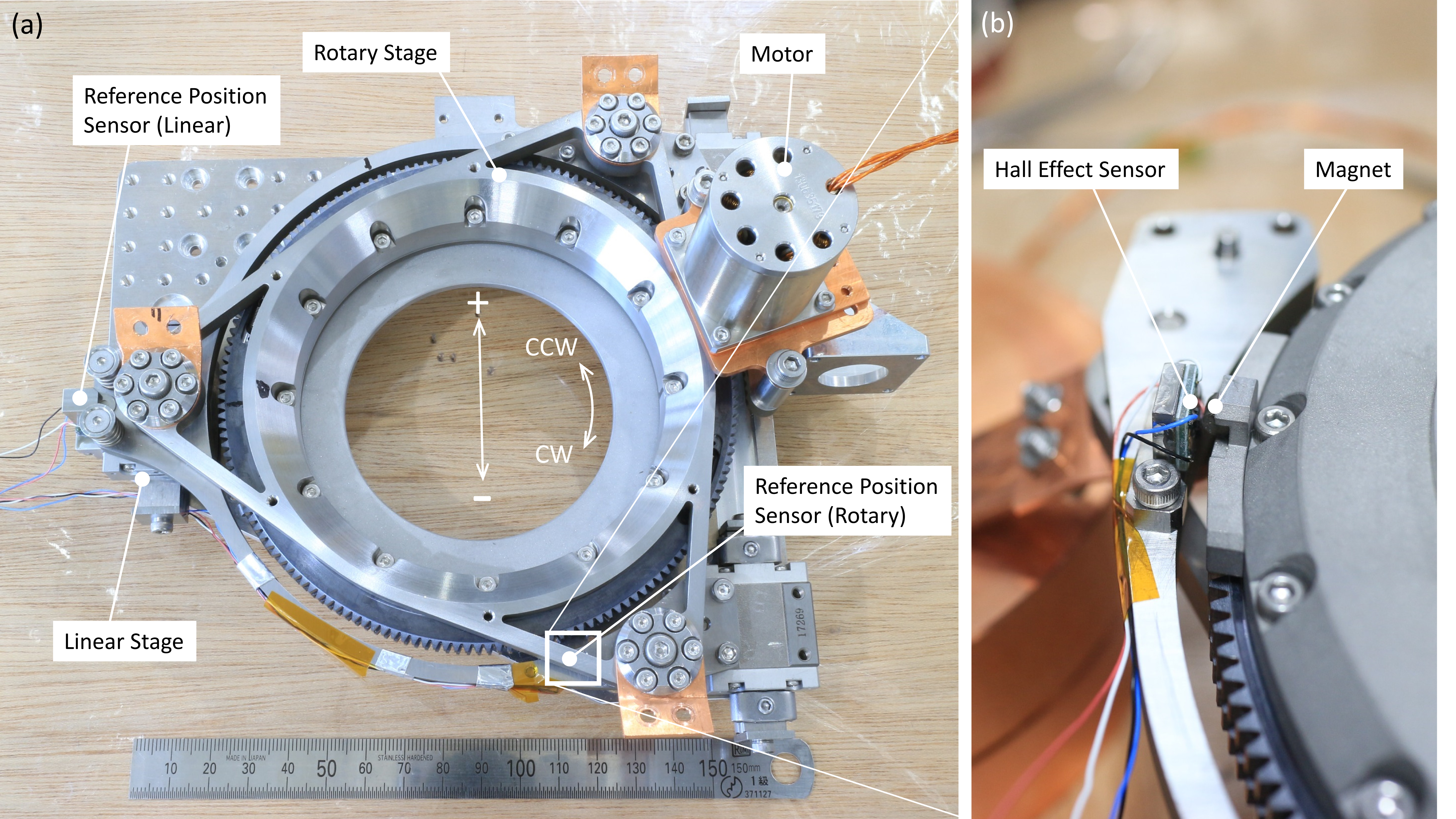}
   \end{tabular}
   \end{center}
   \caption[prototype] 
   { \label{fig:prototype} 
(a) Finished Prototype. The motor and the ball screw for the linear stage are not installed. Directions of rotational and translational motion are defined as shown. (b) Implementation of the hall effect sensor and the magnet for the rotary stage reference position.
}
   \end{figure}

Components for motion control are selected to take advantage of heritage from existing astronomical instruments. Stepper motors from Phytron, Inc. were chosen because of their proven performance under cryogenics. We selected VSS32.200.1,2-UHVC for both rotary and linear stages. The motion controller and motor drivers from Newport Corporation were employed for the prototype. At this combination of the motor and the driver, it was estimated that the motor could generate approximately 30 mNm holding torque at 1.2 A/phase rated supply current. Although the stage is no feedback loop control, the reference position must be defined. The reference position signal is provided by HG-106A hall effect sensor from Asahi Kasei Microdevices Corporation. Fig.~\ref{fig:prototype}(b) shows implementation of the hall effect sensor and the magnet. Because the output of the hall effect sensor is analog, AQV221 Photo MOS relay from Panasonic Corporation is used to convert analog signal to contact signal. 

\section{Test Equipment and Configurations}
\label{sec:testconfiguration}
  
\begin{figure} [ht]
   \begin{center}
   \begin{tabular}{c} 
   \includegraphics[width=14cm]{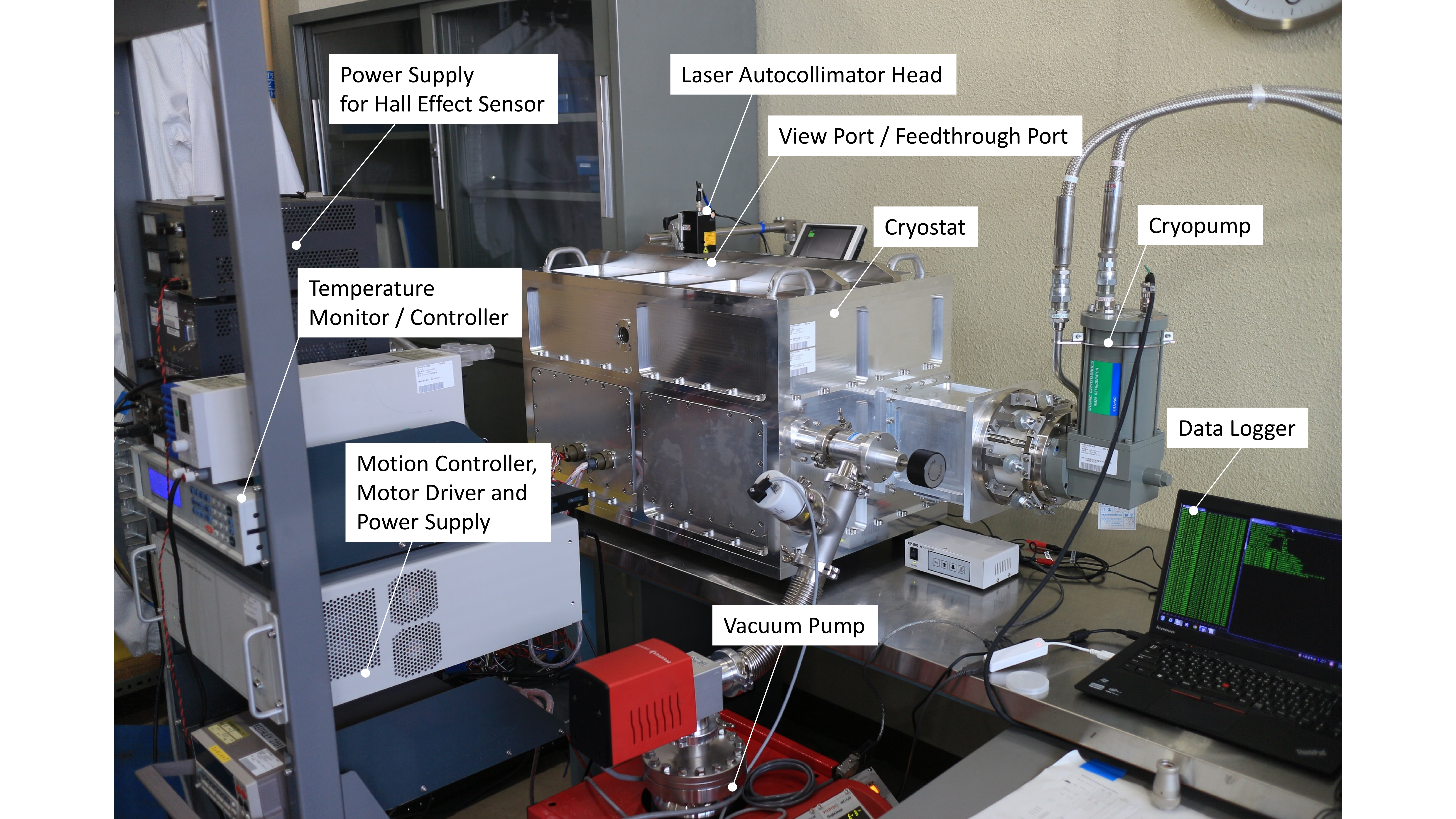}
   \end{tabular}
   \end{center}
   \caption[testequip] 
   { \label{fig:testequip} 
Test equipment. Laser autocollimator is aligned to the top view port of the cryostat.}
   \end{figure}

Test equipment consists of a cryostat, a motion controller, motor drivers and measurement instruments. The cryostat with 30 W cooler and 335 mm (H) $\times$ 350 mm (W)  $\times$  610 mm (D) internal dimension equips 4 ports to enhance the versatility. The ports accommodate several measurement instruments such as a rotary encoder, a torque meter and autocollimator. Fig.~\ref{fig:testequip} shows the test equipment with autocollimator. A temperature controller maintains the stage at 79 $\sim$ 82 K except for the lifetime test that was carried out at 44 $\sim$ 46 K.
 
As described in the 
tab.~\ref{tab:goal}, the prototype will be evaluated in four configurations; wobbling error and drive accuracy in rotary motion, drive accuracy in linear motion and lifetime. The first configuration, for the measurement of wobbling error, one target mirror was located on the center of the rotary stage and we observed its wobble by a laser autocollimator, H350R-C050/F12 from Suruga Seiki co., ltd., through the view port on the top panel. The second configuration, for the measurement of drive accuracy in rotary motion, a drive shaft linked the rotary stage and an angular encoder through ferrofluidic feedthrough in order to compare command angles to actual angles while the stage rotated. The third configuration, for the measurement of drive accuracy in linear motion, a laser displacement sensor, LK-G85 from Keyence corporation, measured and recorded positions of a target piece attached on moving part. One side view port was used for this configuration so that we can leave the laser displacement sensor outside of the cryostat. The last configuration, for the lifetime test, there are several setups for the each component to be evaluated. Because the lifetime test takes very long time, we have accomplished only for the ball bearing as of May 2016. For the ball bearing, the rotary stage without the motor was driven by another stepper motor located outside of the cryostat. A torque meter was inserted into the drive axis to observe fluctuation of drive torque. 


\section{Performance}
\label{sec:performance}

The performance proving test was carried out to evaluate wobbling error, drive accuracy and lifetime which are described in the 
tab.~\ref{tab:goal}. 

\subsection{Rotary Stage}

Tab.~\ref{tab:performance.rotation} summarizes the performance of the rotary stage. 
This result includes 
friction and reaction force of the thermal strap that attached on the rotation part in order to save time for cooling.

\begin{table}[ht]
\caption{Summary of the rotary stage performance. Positioning error in italic is calculated on the assumption that the lost motion is eliminated by feed-forward control.
} 
\label{tab:performance.rotation}
\begin{center}       
\begin{tabular}{lccl} 
\hline
\rule[-1ex]{0pt}{3.5ex}   & Achieved & Goal &  \\ 
\hline
\rule[-1ex]{0pt}{3.5ex}  Wobbling & 0.0022 & 0.05 & $\rm deg_{rms}$   \\
\\
\rule[-1ex]{0pt}{3.5ex}  Positioning error & 0.080 (1.2 A), 0.083 (0.6 A) & 0.05 & $\rm deg_{0-p} (1\sigma)$  \\
\rule[-1ex]{0pt}{3.5ex}  {\small ~~~ Tracking error} & {\small 0.073 (1.2 A), 0.076 (0.6 A)} & {\small N/A} & {\small $\rm deg_{0-p} (1\sigma)$} \\
\rule[-1ex]{0pt}{3.5ex}  {\small ~~~ Lost motion} & {\small 0.057 $\pm$ 0.006 (1.2 A), 0.058 $\pm$ 0.006 (0.6 A)} & {\small N/A} & {\small deg}  \\
\rule[-1ex]{0pt}{3.5ex}  {\small ~~~ Reference position error} & {\small 0.007} & {\small N/A} & {\small $\rm deg_{rms}$}  \\
\rule[-1ex]{0pt}{3.5ex}  {\it Positioning error} & {\it 0.023 (1.2 A), 0.025 (0.6 A)} & {\it 0.05}  & $\it deg_{0-p} (1\sigma)$  \\
\\
\rule[-1ex]{0pt}{3.5ex}  Lifetime (Ball bearing) &  2.3+ & 2.2  & $\times 10^6$ deg  \\
\hline
\end{tabular}
\end{center}
\end{table}

\begin{figure} [ht]
   \begin{center}
   \begin{tabular}{c} 
   \includegraphics[height=8cm]{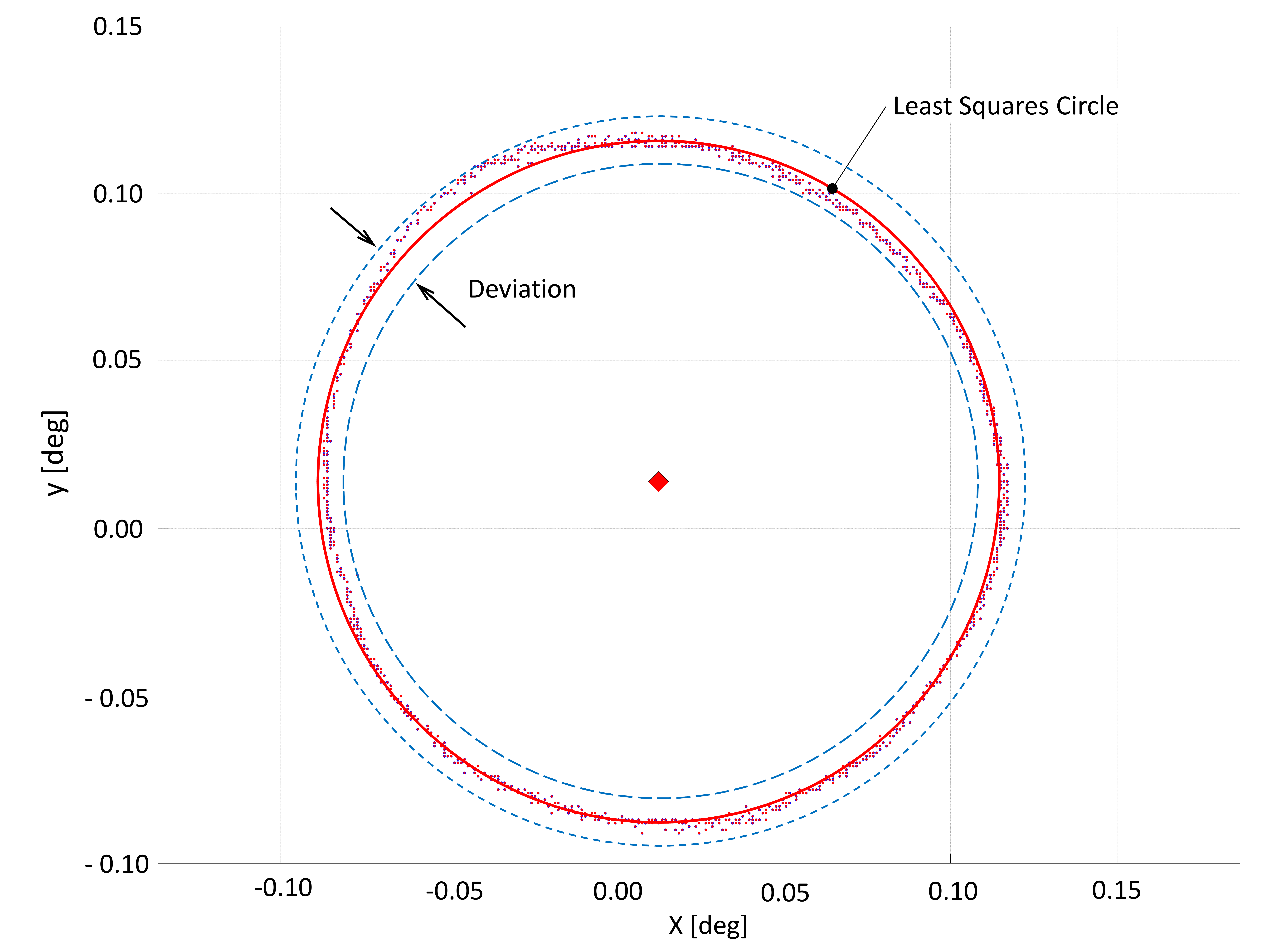}
   \end{tabular}
   \end{center}
   \caption[test] 
   { \label{fig:wobble} 
Trail of the spots (dots) in wobbling error measurement. Deviation from least square circle (red line) corresponds to wobbling error.}
   \end{figure}

First, wobbling error was measured with the laser autocollimator. On 2D detector of the laser autocollimator, the trail of the spots drew a circle such as fig.~\ref{fig:wobble} after one revolution of the stage. This is mainly caused by misalignment of the laser autocollimator axis, the rotation axis and target mirror 
surface normal, which can be corrected or optimized in the case of the final product. Therefore, only the deviation from a least square circle was evaluated. The deviation was obtained as 0.0022 $\rm deg_{rms}$ which meet the development goal. 

\begin{figure} [ht]
   \begin{center}
   \begin{tabular}{c} 
   \includegraphics[height=6cm]{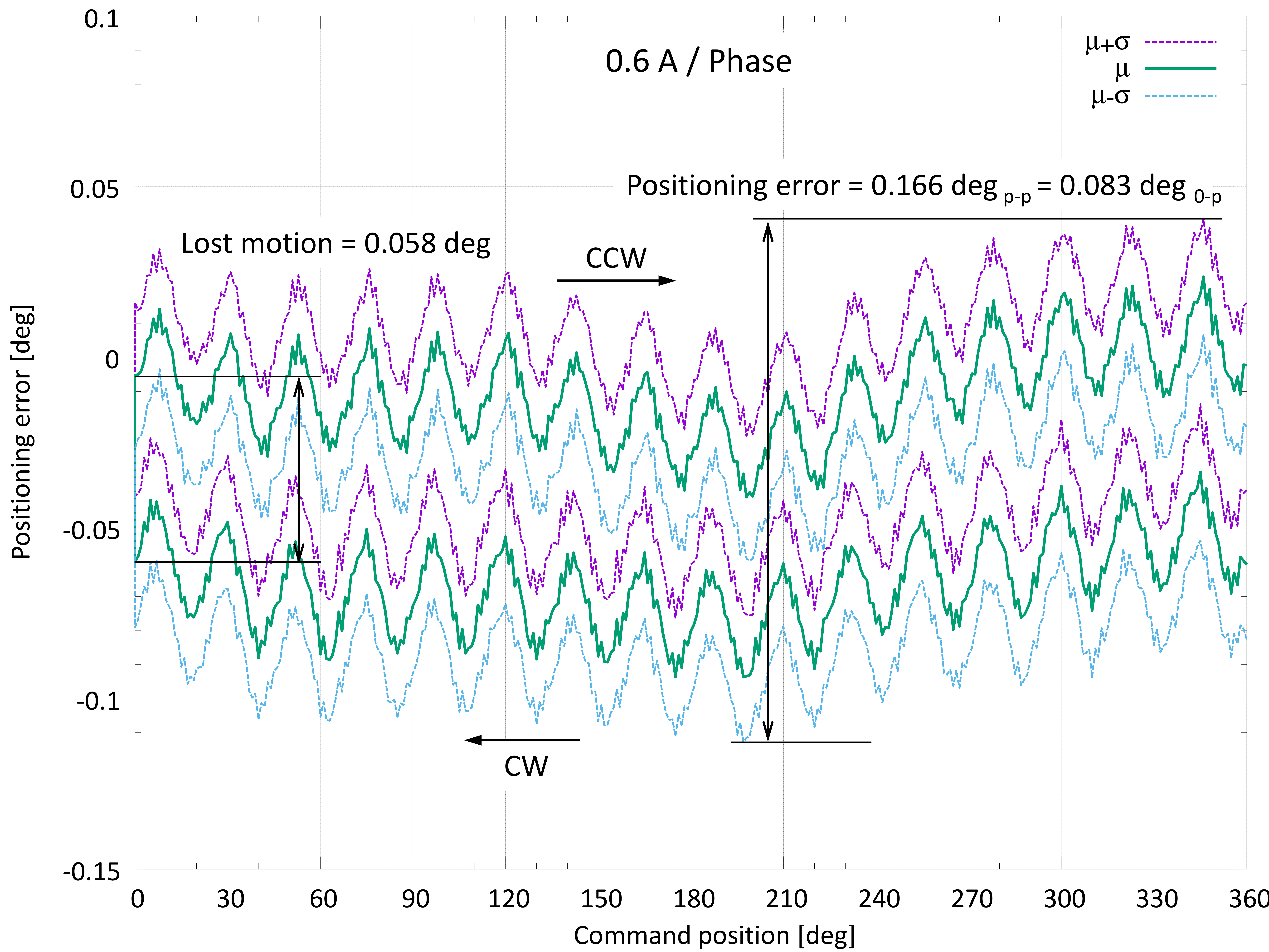}
   \includegraphics[height=6cm]{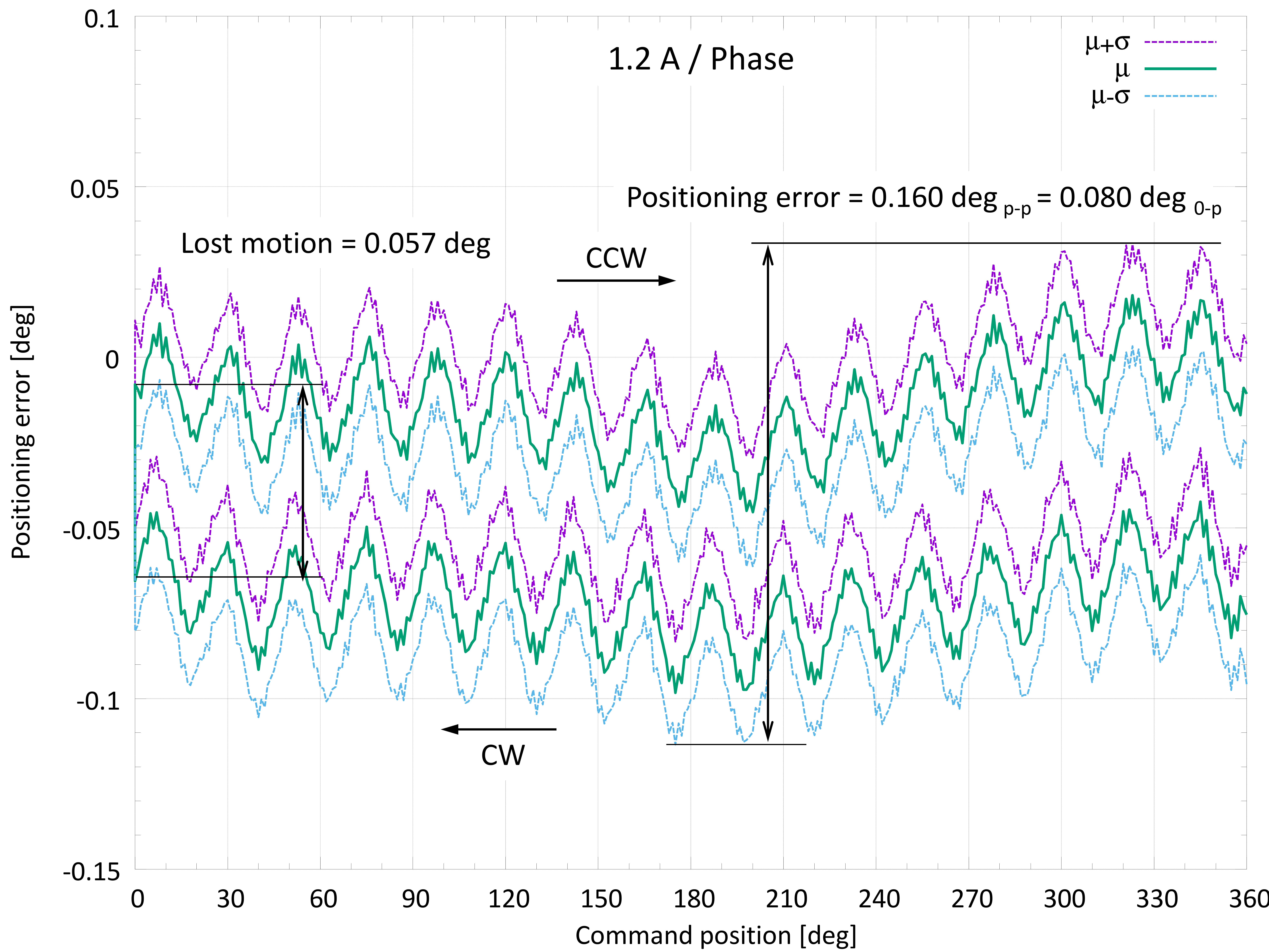}
   \end{tabular}
   \end{center}
   \caption[test] 
   { \label{fig:rot} 
The rotary positioning error of 0.6 A (left) and 1.2 A (right).
The mean waveform $\mu$ and 1$\sigma$ band are calculated by three independent measurements.
}
   \end{figure}

Next, the tracking error during one revolution in CW and CCW directions was measured as the part of drive accuracy of rotary motion. Because the tracking error will increase when the input current for the motor is reduced, the measurement was carried out with two input current settings; 1.2 A per phase (100\% rated current) and  0.6 A per phase (50 \% rated current). As expected, 0.6 A setting shows slightly worse performance than 1.2 A setting. Fig.~\ref{fig:rot} shows the tracking error waveforms of two input current configurations. In the waveform, two notable features were observed. First, there is cyclic pattern whose frequency was sixteen cycles per revolution. It is considered that these cyclic pattern is caused by eccentricity of gears because sixteen cycles per revolution corresponds to gear ratio of drive and driven gear. Second, there is a large offset between CW and CCW direction. This is so called the lost motion that seems to be combination of mechanical backlash and deadband of the stepper motor. The total positioning error was calculated as a combination of the tracking error and the reference position error. The reference position error was 0.007 $\rm deg_{rms}$ which does not give a crucial impact in the total positioning error of 0.080 $\rm deg_{0-p}$ for 1.2 A setting, for example.

Under the present results, the performance does not reach the development goal even though 100 \% rated current was supplied. So we are going to implement feed-forward control to reduce the total positioning error.
The feed-forward control can be applicable when the error is predictable. As fig.~\ref{fig:rot} shows, the systematic component of the lost motion will be predictable because the waveforms of CW and CCW seem to be simply offset. This offset can be measured through a referencing sequence. For example, the rotary stage rotates to find a reference position in CW direction, then rotates in opposite CCW direction for one revolution. In CCW motion, the rotary stage will not detect the reference position signal because the travel of the rotary stage will have shortage due to the lost motion. Finally, additional travel in CCW to detect the reference position will correspond to the lost motion. The high level controller can be capable to implement this bi-directional referencing sequence and add an appropriate offset on the position command in the normal operation. If this strategy worked as expected, the tracking error of 1.2 A setting could be reduced from 0.080 $\rm deg_{0-p}$ to 0.023 $\rm deg_{0-p}$ by subtracting systematic component of the lost motion. It is to be noted that this expectation might be optimistic because the calibration error is not included.
 
\begin{figure} [ht]
   \begin{center}
   \begin{tabular}{c} 
   \includegraphics[height=8cm]{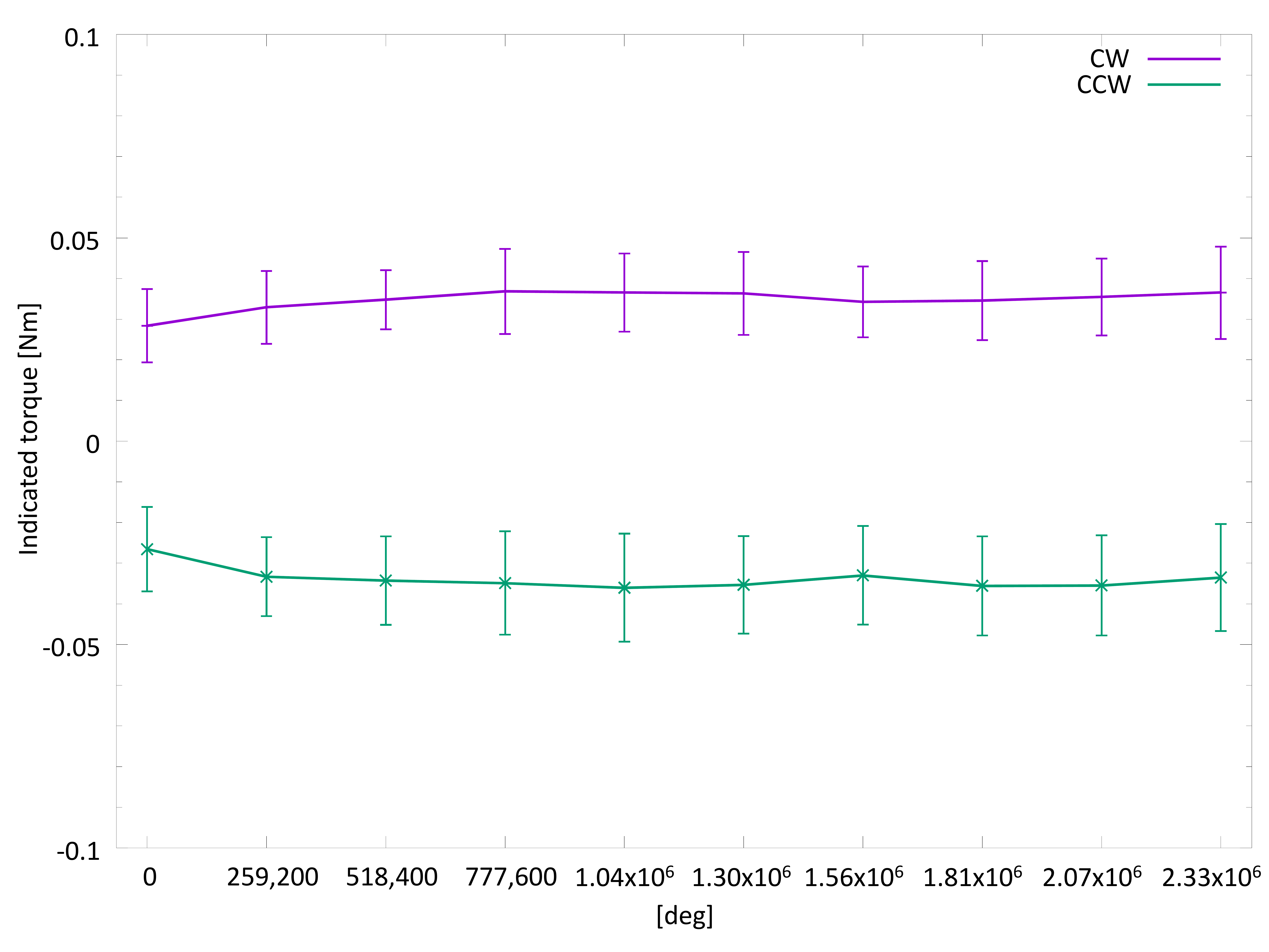}
   \end{tabular}
   \end{center}
   \caption[test] 
   { \label{fig:lifetime} 
Torque variation of the rotary stage during the durability test. Torque in CW and CCW rotation are plotted separately. Error bars represent standard deviation in one revolution. }
   \end{figure}

At the last, the lifetime test of the ball bearing was carried out after all other tests completed. We put 3.4 kg payload on the stage and drove 2.3 $\times 10^6$ deg in total. The rotation torque during one revolution in 
CW and CCW directions were recorded each 2.5 $\times 10^5$ deg. Because the variation of the torque shown in fig.~\ref{fig:lifetime} looks steady, no significant degradation will not occur within estimated lifetime period. Another remaining item to be evaluated is lifetime of drive and driven gears. We are planning to carry out the gear lifetime test in 2016.

\subsection{Linear Stage}

\begin{figure} [ht]
   \begin{center}
   \begin{tabular}{c} 
   \includegraphics[height=6cm]{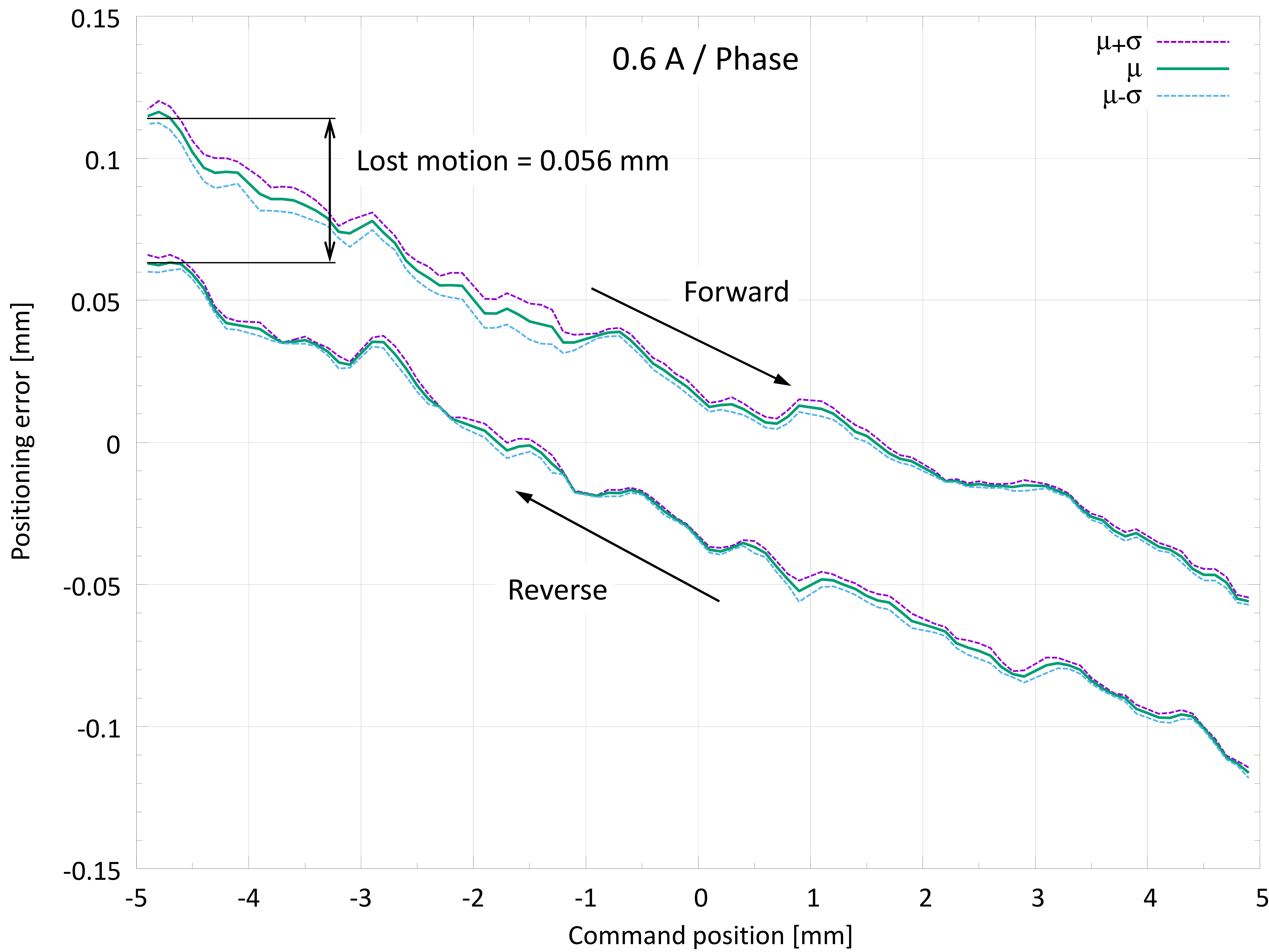}
   \includegraphics[height=6cm]{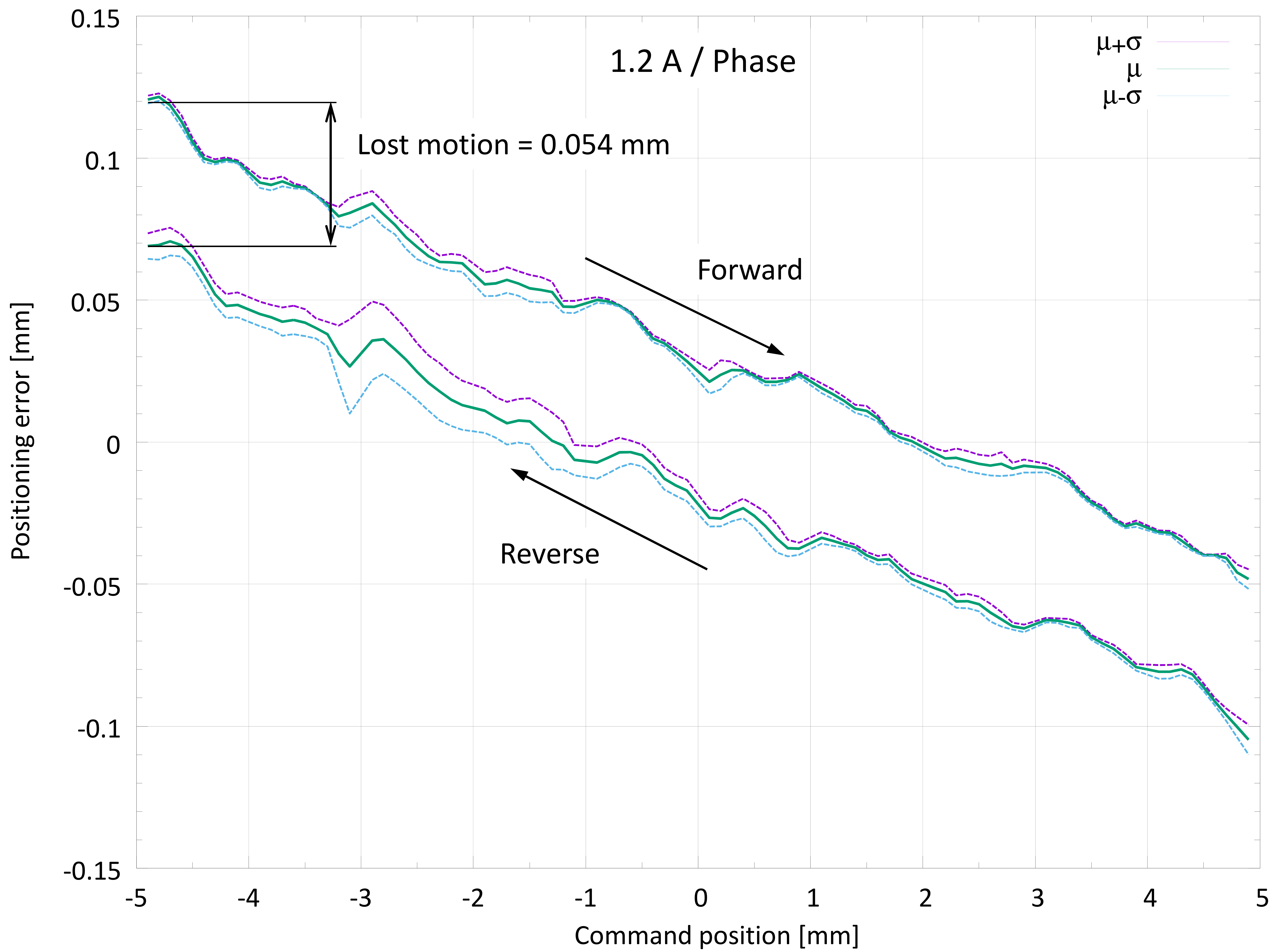}
   \end{tabular}
   \end{center}
   \caption[test] 
   { \label{fig:linear} 
The linear positioning error of 0.6 A (left) and 1.2 A (right).
The mean waveform $\mu$ and 1$\sigma$ band are calculated by three independent measurements.
}
   \end{figure}

Tab.~\ref{tab:performance.linear} summarizes the performance of the linear stage.  First, the reference position error was measured and it turned out that this error was 0.001 mm which is sufficiently small. Next, the positioning error was measured along with $\pm$ 5 mm round-trip by using the 
laser displacement sensor. As well as the rotary stage, the measurement was carried out with two input current settings; 1.2 A per phase (100\% rated current) and  0.6 A per phase (50 \% rated current). The result is shown in fig.~\ref{fig:linear}. The plots have a constant slope that suggests a major component of the error is in linearity. This is similar trend as known as a lead accuracy of a ball screw, however the deviation of 
0.23 $\sim$ 0.24/$\pm$ 5 mm looks too large as compared with the original specification of the ball screw that is $\pm$ 0.018 mm/42.5 mm. The lost motion of 0.05 $\sim$ 0.06 mm is also large. 

\begin{table}[ht]
\caption{Summary of the linear stage performance. Deviation is scaled to be consistent with the positoning error in the development goal. Positioning error in italic is calculated on the assumption that the linearity is compensated and the lost motion is eliminated by feed-forward control.} 
\label{tab:performance.linear}
\begin{center}       
\begin{tabular}{lccl} 
\hline
\rule[-1ex]{0pt}{3.5ex}   & Achieved & Goal &  \\ 
\hline
\rule[-1ex]{0pt}{3.5ex}  Positioning error & 0.070 (1.2 A), 0.072 (0.6 A) & 0.05 & $\rm mm_{0-p}/\pm 3mm (1\sigma)$  \\ 
\rule[-1ex]{0pt}{3.5ex}  {\small ~~~ Reference position error} & {\small 0.001} & {\small N/A} & {\small $\rm  mm_{rms}$}  \\
\rule[-1ex]{0pt}{3.5ex}  {\small ~~~ Deviation} & {\small 0.116 (1.2 A), 0.120 (0.6 A)} &  {\small N/A}  & {\small $\rm mm_{0-p}/\pm5 mm (1\sigma)$}  \\
\rule[-1ex]{0pt}{3.5ex}  {\small ~~~ Lost motion}  & {\small 0.054 $\pm$ 0.004 (1.2 A), 0.056 $\pm$ 0.005 (0.6 A)} &  {\small N/A}  & {\small mm}  \\
\rule[-1ex]{0pt}{3.5ex}  {\it Positioning error}  & {\it 0.004 (1.2 A), 0.005 (0.6 A)} & {\it 0.05} & $\it mm_{0-p}/\pm3 mm (1\sigma)$   \\
\\
\rule[-1ex]{0pt}{3.5ex}  Lifetime (Ball screw) &  7.3+ (Non coated), 2.0+ (MoST\texttrademark), & 45 & $\times 10^5$ mm  \\
\rule[-1ex]{0pt}{3.5ex}  &  1.7+ (Silicon nitride) &  &   \\
\hline
\end{tabular}
\end{center}
\end{table}

There is no doubt that the current performance underreaches the development goal. Increase of the input current for the motor could not improve the performance significantly, so we have to implement feed-forward control as well as the rotary stage.  As for the feed-forward control, this will be similar to one for the rotary stage, however not only the lost motion but also the linearity error must be corrected. 
At least, three reference position sensors at the center and both ends of stroke are required to compensate the lost motion and the linearity error. The positions of these reference sensors must be measured by another reference such as a laser displacement sensor to know actual positions. In the referencing sequence, the stage will move one round trip to scan all three sensors bidirectionally, and then the high level controller will calculate regression coefficient and offset of the position error curve by comparing command positions with a record of actual positions. Also the high level controller will make correction table and interpolate to generate appropriate position command. Residue of linearity compensation and lost motion elimination 
are added in tab.~\ref{tab:performance.linear} as an expected performance.

The lifetime test of the linear stage components have been carried out since Feb 2016. The first component is the ball screw. We prepared three samples that have different kind of balls. The ball screws with non-coated, MoST\texttrademark coating and silicon nitride balls have survived at 7.3 $\times 10^5$ mm, 2.0 $\times 10^5$ mm and 1.7 $\times 10^5$ mm respectively with 20 N axial force, and they are still functional as of May 2016.  We will continue the lifetime test until they reach the development goal or have failures. The next component will be the linear motion guide. Its lifetime test has also been prepared and will be started after the ball screw durability test in 2016.

\section{Conclusion}

We completed mechanical design and fabrication of a two axis stage prototype with the immediate development goal probable for the stages of the IRIS Imager. The performance proving tests have been carried out to evaluate errors such as wobbling, rotary and linear positioning, and lifetime. At present results, positioning errors in rotary and linear motion 
achieve the goal with conditions such that 
the errors are reduced by implementing feed-forward control to eliminate systematic component of the error. 

\acknowledgments 

The TMT Project gratefully acknowledges the support of the TMT collaborating institutions.  They are the California Institute of Technology, the University of California, the National Astronomical Observatory of Japan, the National Astronomical Observatories of China and their consortium partners, the Department of Science and Technology of India and their supported institutes, and the National Research Council of Canada.  This work was supported as well by the Gordon and Betty Moore Foundation, the Canada Foundation for Innovation, the Ontario Ministry of Research and Innovation, the Natural Sciences and Engineering Research Council of Canada, the British Columbia Knowledge Development Fund, the Association of Canadian Universities for Research in Astronomy (ACURA) , the Association of Universities for Research in Astronomy (AURA), the U.S. National Science Foundation, the National Institutes of Natural Sciences of Japan, and the Department of Atomic Energy of India. We would like to thank Mr. Mitsuhiro Fukushima, Mr. Takeo Fukuda, Mr. Hikaru Iwashira, Mr. Tetsuo Nishino and Mr. Norio Okada of Mechanical Engineering Shop for their experienced craftsmanship on manufacturing of the prototype. We are particularly grateful for the support given by Dr. Hiroshi Matsuo of ATC in establishment and improvement of our experiment environment. Mr. Takashi Nakamoto of NAOJ gives insightful comments and suggestions for the performance proving test.

\bibliography{report} 
\bibliographystyle{spiebib} 
\end{document}